\documentclass[lettersize,journal]{IEEEtran}
\usepackage{amsmath,amsfonts}
\usepackage{mathtools}
\usepackage{algorithmic}
\usepackage{algorithm}
\usepackage{array}
\usepackage[caption=false,font=normalsize,labelfont=sf,textfont=sf]{subfig}
\usepackage{textcomp}
\usepackage{stfloats}
\usepackage[hyphens]{url}
\usepackage{verbatim}
\usepackage{graphicx}
\usepackage{cite}
\usepackage{booktabs} % for professional tables
\usepackage{adjustbox, diagbox, adjustbox, caption}
\usepackage{tabularx, multirow, multicol, makecell}
\begin{document}

\title{Leveraging Spatial Cues from Cochlear Implant Microphones to Efficiently Enhance Speech Separation in Real-World Listening Scenes}

\author{
  \IEEEauthorblockN{Feyisayo Olalere\textsuperscript{1},
  Kiki van der Heijden\textsuperscript{1,2},
  Christiaan H Stronks\textsuperscript{3,4},
  Jeroen Briaire\textsuperscript{3},
  \textit{Senior Member, IEEE}, Johan HM Frijns\textsuperscript{3,4,5},
  \textit{Senior Member, IEEE}, Marcel van Gerven\textsuperscript{1}}\\
  \IEEEauthorblockA{
    \textsuperscript{1}Radboud University, The Netherlands \\
    \textsuperscript{2}Mortimer B. Zuckerman Mind, Brain, Behavior Institute, Columbia University, USA \\
    \textsuperscript{3}Department of Otorhinolaryngology, Leiden University Medical Centre, The Netherlands \\
    \textsuperscript{4}Leiden Institute for Brain and Cognition, Leiden, The Netherlands\\
    \textsuperscript{5}Department of Bioelectronics, Delf University of Technology, Delft, The Netherlands
  }
}

\maketitle

\begin{abstract}
Speech separation approaches for single-channel, dry speech mixtures have improved greatly. However, speech separation in real-world, spatial and reverberant acoustic environments remains challenging. This limits the efficiency of existing approaches for real-world speech separation applications in assistive hearing devices such as cochlear implants (CIs). To address this issue, we quantify the impact of real-world acoustic scenes on speech separation and investigate to what extent spatial cues from such real-world scenes can improve separation quality in an efficient manner. Crucially, we characterize speech separation performance as a function of implicit spatial cues (i.e., cues inherently present in the acoustic input that can be learned by the model), as well as of explicit spatial cues (i.e., manually calculated spatial features added as auxiliary input to the model). Our findings show that spatial cues (both implicit and explicit) improve separation performance for mixtures with spatially separated talkers, but also for mixtures with nearby talkers. Further, we demonstrate that spatial cues enhance speech separation in particular when spectral cues for separation are ambiguous, that is, when voices are similar. Finally, we show that the addition of explicit, auxiliary spatial cues is particularly beneficial when implicit spatial cues are weak. For example, microphone recordings from a single CI contain weaker implicit spatial cues than when microphone recordings from two, bilateral CIs are combined. These findings emphasize the importance of training models on real-world data to improve generalisability to everyday listening situations and contribute to the development of more efficient speech separation approaches for CIs or other assistive hearing devices in such real-world listening situations.
\end{abstract}


\section{Introduction}
\label{sec:intro}

\IEEEPARstart{H}{earing} impairments affect more than 5\% of the global population and result in significant communication difficulties, even with the use of assistive hearing devices such as cochlear implants (CIs)\cite{who2019,caldwell2013speech}. CIs are implanted neuroprosthetic devices which directly stimulate the auditory nerve to restore hearing for individuals with severe-to-profound sensorineural hearing loss\cite{naples2020cochlear}. While CI users experience tremendous improvements in speech perception in quiet environments, communication difficulties persist in noisy, everyday listening scenes such as the classroom, the office, or at social gatherings\cite{qin2003effects,stickney2004cochlear,healy2016difficulty,shinn2008selective}. These difficulties are a consequence of peripheral processing deficits, which hamper a CI user's ability so selectively attend to a talker in noisy listening scenes \cite{shinn2008selective}.

Implementing a speech denoising algorithm as a front-end processing step in a CI or other assistive hearing device significantly improves speech perception for hearing impaired listeners, because this provides the user with clean speech signals. However, everyday listening environments often consist of multiple, simultaneous talkers in addition to various background noise sources \cite{rowland2018listening}. As voices have similar acoustic characteristics, such multi-talker scenes are particularly challenging for speech denoising algorithms. Therefore, automatic speech separation also plays a crucial role in front-end speech processing \cite{kokkinakis2008using}.

Currently, deep neural network (DNN) based approaches are the state of the art for speech separation. Time-frequency based DNN speech separation methods operate on spectrogram representations of multi-talker mixtures\cite{wang2014training,liu2019divide,liu2020causal}. Such approaches either extract masks of each talker in the mixture \cite{williamson2015complex,narayanan2013ideal} or directly reconstruct the spectrogram of each talker\cite{lu2013speech,xu2013experimental}. However, time-frequency speech separation approaches have several drawbacks. Most notably, as phase information is discarded, only an approximated phase\cite{perraudin2013fast} or the phase of the mixed audio\cite{gu2019neural} can be used to reconstruct the sound wave for each talker. This introduces additional noise and limits the quality of the reconstructed, clean talkers \cite{wang2018supervised}.

More recent speech separation approaches therefore focus on separation in the time-domain, i.e., operating directly on the audio mixture's waveform\cite{huang2014deep,zhang2016deep,luo2019conv}. These approaches avoid the phase inversion problem and therefore introduce fewer artifacts in the reconstructed speech waveforms \cite{luo2019conv}. Moreover, several time-domain approaches were shown to be highly computationally efficient \cite{luo2019conv,luo2020dual,tzinis2022compute}, making them more suitable for small devices such as CIs and other assistive hearing technology. 

However, few approaches (either in the time or in the time-frequency domain) have been applied to real-world acoustic scenes that are spatial and include reverberation. Reverberation distorts the talkers' speech and thereby increases the difficulty of the speech separation task significantly \cite{luo2020dual}, resulting in a drop in speech separation performance\cite{gu2020temporal,maciejewski2020whamr,zhao2023mossformer2}. This poses a significant obstacle to the application of speech separation algorithms as a front-end processing step in assistive hearing technology such as CIs. Yet, the spatial dimension of real-world acoustic scenes can also be considered an additional source of information for optimizing speech separation: The spatial separation between talkers causes each talker to have different spatial cues, which can be utilised by speech separation algorithms for speech separation. 

Studies investigating to what degree incorporating spatial learning improves speech separation performance typically add spatial features as auxiliary input to the model. For example, inter-microphone phase differences (IPD) \textit{a priori} \cite{chen2018multi,wang2018combining,gu2019neural,gu2020enhancing}, level differences (ILD) \cite{gu20213d,han2020real}, time difference (ITD) \cite{zohourian2016binaural}, or angle features (AF) \cite{chen2018multi}. The aforementioned approaches all show that auxiliary spatial features improve speech separation performance. However, these spatial features are typically extracted from time-frequency representations of the two-talker mixture, even if the model itself operates in the time domain (e.g.\cite{gu2020enhancing,han2020real}). This results in relatively long latency and a high computational cost. Moreover, introducing auxiliary features such as spatial features increases the number of model parameters, resulting in an even higher latency and computational cost. These approaches are thus sub-optimal for applications in devices such as CIs, which require energy-efficient and low-latency speech separation approaches. Furthermore, a recent study which introduced an efficient speech separation system utilising the spatial information that is present in the scene rather than adding explicit, auxiliary spatial features, showed a significant drop in performance in real-world scenes with reverberation \cite{han2020real}. A similar study \cite{gaultier2024recovering} which also leveraged implicit cues for a single speaker and noise configuration shows the drop in performance experienced in the real-word even with the help of implicit cues.

The present work therefore investigates to what extent the spatial cues directly available in the audio signals recorded by microphones on an assistive hearing device such as a CI can be utilised to efficiently enhance speech separation in real-world listening scenes. We assess the impact of such implicit spatial cues (that is, cues inherently present in the acoustic input) on speech separation and we evaluate how this relates to separation performance when intermicrophone phase differences (IPDs) are added as explicit, auxilliary spatial cues to the model input. Further, as CI implantation practices vary globally, we quantify speech separation performance gains both for use cases in which a listener is implanted with a unilateral CI as well as use cases in which a listener is implanted with bilateral CIs. Furthermore, we analyzed the effect of the cues on the angle of separation between the speakers. As well as the effect of the perceived gender on the separation performance. For these experiments, we make use of an adapted version of the SuDoRM-RF model, which is a highly efficient time-domain approach with state-of-the-art speech separation performance \cite{tzinis2022compute}.  

Our results illustrate the detrimental effect of real-world acoustic environments on speech separation performance and demonstrate that spatial cues (either implicit or explicit) substantially improves speech separation. Crucially, our findings characterize the challenges of efficient, low-latency speech separation in real-world, ecologically valid acoustic scenes and contribute towards the development of more effective strategies to enhance speech separation performance for front-end speech processing in CIs and other assistive hearing devices. 

The remainder of the paper is organized as follows: Section II describes the task of speech separation, the generation of the real-world, spatialized dataset and the model architecture. We describe the experimental framework for this study in Section III. In Section IV, we present the results and analysis. Section V discusses the implications of our findings for speech enhancement in real-world listening scenes and CIs in particular. We conclude the paper in Section VI.

\section{Speech separation in real-world scenes}

\subsection{Task}
\noindent In this study, the task of speech separation consists of estimating the waveform of talker $s_1$ and $s_2$ from the discrete waveform of the speech mixture \textbf{y} $\in \mathbb{R}^{C \times T}$. Here, $C$ represents the number of channels (i.e., microphones) and $T$ denotes time. 

\subsection{Dataset}
\noindent We simulated a dataset consisting of two-talker mixtures in spatial, real-world listening scenes including reverberation usingthe WSJ0-2mix dataset \cite{hershey2016deep} and a custom sound spatialization pipeline. The WSJ0-2 dataset is extensively employed for speech separation tasks (e.g., \cite{huang2014deep,zhang2016deep,luo2019conv}) and comprises 20,000 training, 5,000 validation and 3,000 test samples of two-talker mixtures.  

\begin{figure}[!t]
\includegraphics[width=\linewidth]{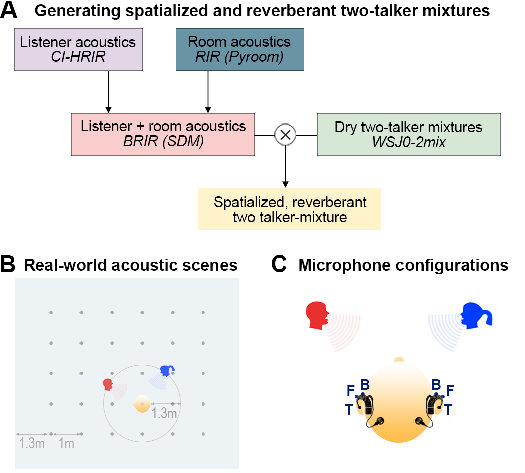}
\caption{(A) Schematic depiction of data generation. CI-HRIR = Cochlear implants Head Impulse Response; RIR = Room impulse response; BRIR = Binaural Room Impulse Response. (B) An acoustic scene depicting talkers positioned around a listener located in a room. (C) Schematic example of an acoustic scene with two talkers and a listener wearing bilateral CIs. Each CI has three microphones: F = front microphone, B = back microphone and T = T-microphone.}
\label{img:spatialisation}
\end{figure}

The spatialization pipeline consisted of four components (Fig. \ref{img:spatialisation}): (1) Simulation of room impulse responses (RIRs), (2) CI-HRTFs, (3)
Generation of binaural room impulse responses (BRIRs); (4)  Mixing real-world two-talker mixtures.

\subsubsection{Room Impulse Response (RIR)}
\label{rir}
A RIR describes room-specific acoustic properties including reverberation, reflection, and echo. Here, we used the Pyroomacoustics Python package \cite{scheibler2018pyroomacoustics}, which  employs an image source model to efficiently simulate RIRs by simulating sound wave propagation from a source to a receiver within a shoebox room. In total, we simulated 500 shoebox rooms with different dimensions and reverberation properties (Fig. \ref{img:spatialisation} A). Room dimensions were randomly selected from a range of $4 \times 4 \times 2.5$ m to $10 \times 10 \times 5$ m  (length $\times$ width $\times$ height), encompassing common sizes of classrooms, meeting rooms, and restaurants \cite{commercial_acoustics}. 

\begin{figure*}[ht]
    \centering
    \includegraphics[width=\linewidth]{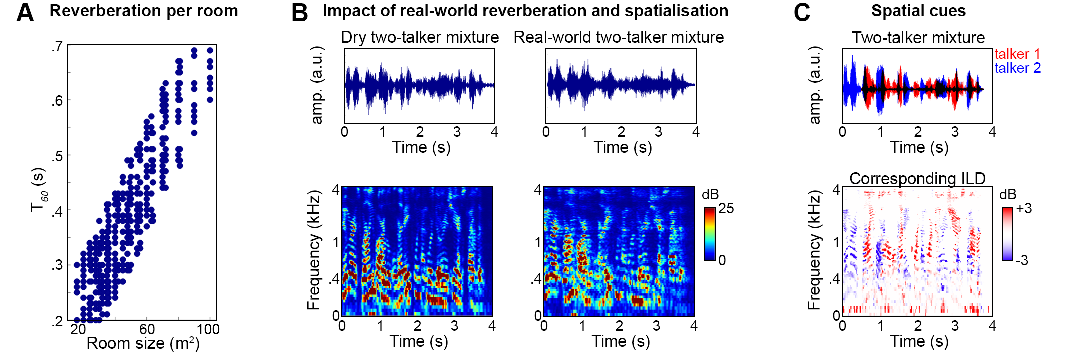}
    \caption{(A) Reverberation time ($T_{60}$) as a function of room size. Each circle represents a single room. (B) Effect of reverberation and spatialization on speech mixtures. Top panes show waveforms of a dry, non-spatial speech mixture and a spatial, reverberant version of the same speech mixture. For illustration, bottom panes depict spectrograms (but note that models are trained directly on the waveform). (C) Presence of implicit spatial cues in real-world acoustic scenes. Top pane shows an example of a two-talker mixture in a spatial, reverberant scene. One talker is at -90$^{\circ}$, the other talker is at +90$^{\circ}$. Bottom pane visualizes the corresponding interaural level differences for this speech mixture.}
    \label{img:rt60}
\end{figure*}

For each Room Impulse Response (RIR), we sampled reverberation time ($T_{60}$) from a range of 0.2 s to 0.7 s (stepsize =  0.01 s). The reverberation time reflects the strength of reverberation in a room and is dependent both on the size of the room and the materials of which the room consists \cite{larson_davis,acoustic_frontiers}. To introduce a naturalistic relation between room size and $T_{60}$, we restricted the range of $T_{60}$ to sample from based on room size. That is, we scaled all room sizes used in the present study between 0 and 1 and used this factor as an index to sample $T_{60}$ from the range of 0.2 to 0.7 s. The resulting distribution of $T_{60}$ as a function of room size is depicted in Fig. \ref{img:rt60}A. Additionally, to introduce variability, we added a random offset within a range of -10 to 10 ms to the selected reverberation time. 

To define the position of a listener within the room, we positioned a grid of potential listener locations (that is, receiver locations) at the center of the room with axes running parallel to the walls (1.4 m distance from the wall). Potential listener locations were spaced 1 m apart. We randomly sampled five locations from the grid for each room but the smallest room to simulate five listener locations per room. The smallest room accommodated only four listener locations. 

We simulated RIRs for a microphone geometry composed of six DPA-4060 omnidirectional microphones arranged in orthogonal pairs and a central Earthworks M30/M50 microphone. This geometry (FRL-10) with a diameter of 10 cm was used to ensure correspondence with the subsequent binaural rendering method used here \cite{amengual2021optimizations} (Section \ref{BRIR} Binaural Room Impulse Response).

To define the position of the talker (i.e., the source) with respect to the listener, we positioned a circle (radius = 1.4 m) around the selected listener location. On this circle, we indicated 24 talker (source) locations in equidistant steps of 15$^{\circ}$.  Both talker locations and listener locations were positioned at a height of 1.25 m, simulating the seated position of the listener and talkers.  This procedure led to the generation of a total of 59,688 RIRs.

\subsubsection{Cochlear Implant Head Related Impulse Response}
\label{ci-hrtf}
\noindent Listener-specific acoustic properties are captured by a head-related impulse response (HRIR). That is, the HRIR reflects the impact of the pinnae (outer ears), head, and torso on sound waves arriving at the ear canal\cite{gardner1994hrft}. Here, we made use of HRIRs recorded with the microphones on a CI which were provided for this study by CI manufacturer Advanced Bionics (www.advancedbionics.com), resulting in ecologically valid, CI-specific listener acoustic attributes (see Fig. \ref{img:spatialisation}B). 

The CI-HRIRs were measured by fitting the left and right ear of a KEMAR mannequin with a CI with three microphones each (Fig. \ref{img:spatialisation}B). 
The behind-the-ear (BTE) microphones (Front and Back mic), positioned at the back and front of the device, are known to be more susceptible to environmental noise \cite{jones2016effect, kolberg2015cochlear}. Conversely, the T-mic \cite{t-mic}, situated at the entrance to the ear canal is less susceptible to noise and captures spatial cues similar to those captured by the ears of normal hearing listeners\cite{mayo2020acoustic}. We therefore selected the T-mic as model input for input configurations consisting of single channel per device and added the back-mic for input configurations consisting of two channels per device.  
Finally, CI-HRIRs were recorded for 24 azimuth locations (from 0$^{\circ}$ to 345$^{\circ}$ in 15$^{\circ}$ increments) at 0$^{\circ}$ elevation.

\subsubsection{Binaural Room Impulse Response (BRIR)}
\label{BRIR}
\noindent The BRIR combines the room-specific acoustic properties (i.e., the RIR) and listener-specific acoustic properties (i.e., the CI-HRIR) \cite{gari2019flexible}. Convolving the BRIR with a monaural sound clip therefore results in a spatialized and reverberant sound scene that captures both room-specific and listener-specific acoustic properties. To generate BRIRs from our set of RIRs and the CI-HRIRs, we leveraged the BRIR generator proposed by Amengual et al. \cite{amengual2021optimizations}. which uses the spatial decomposition method proposed in \cite{tervo2013spatial}. This resulted in a set of 59,688 BRIRs. 

\subsubsection{Real-world, two-talker speech mixtures}
We simulated real-world, two-talker mixtures with varying separation angles between talkers: 0\textdegree{}, 15\textdegree{}, 30\textdegree{}, 60\textdegree{} and 90\textdegree{}, corresponding to 19.9\%, 19.98\%, 19.81\%, 20.14\% and 20.18\% of the data, respectively. To this end, we convolved the single-channel waveform of each talker selected for the two-talker mixture (i.e., selected from the WSJ0 dataset) with the \textit{a priori} generated BRIR corresponding to a given acoustic room and talker location. We then summed the spatialized and reverberant waveforms of both talkers to render the two-talker mixtures. Figure~\ref{img:rt60} C shows that the generated sound clips contain realistic spatial cues comparable to typical human spatial cues \cite{oldfield1984acuity, makous1990two}. 

\subsection{Model}

\begin{figure}[ht]
    \centering
    \includegraphics[width=\linewidth]{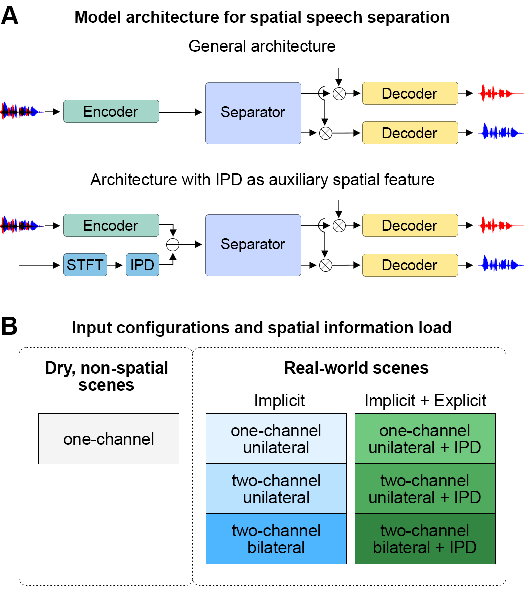}
    \caption{(A) SuDoRM-RF \cite{tzinis2022compute} implementation in present study. Top row depicts SuDoRM-RF approach for input configurations consisting only of the speech mixture (i.e. waveform). Bottom row depicts our SuDoRM-RF approach for input configurations consisting of speech mixture (i.e. waveform) and IPD as an auxiliary feature.  (B) Schematic overview of all input configurations. Color indicates the presence of implicit (blue) and combined implicit and explicit (green) spatial cues. A higher saturation signifies stronger spatial cues.}
\label{img:AdaptedModel}
\end{figure}

\noindent In this study, we adapted the SuDoRM-RF model \cite{tzinis2022compute} to run our various experiments (see Fig \ref{img:AdaptedModel}A). The SudoRM-RF model is highly efficient in terms of computation and memory \cite{tzinis2022compute}, crucial characteristics for models intended to operate on compact devices like CIs \cite{wasmann2021computational}.  

SuDoRM-RF is a time-domain model that consists of an encoder-decoder architecture with three stages: an encoder, a separator, and a decoder. For a comprehensive understanding of these stages, refer to \cite{tzinis2022compute}. In short, the encoder processes the two-talker mixture waveform $y$ through a Conv1D block, generating a latent representation ($R_l$) of the mixture. Subsequently, the separator block learns masks $(M)$ for each talker present in the input mixture ($y$). The learned masks are multiplied with the latent representation of the mixture ($M \times R_l$) to extract a latent representation of each talker. Finally, the decoder block converts the estimated talker's representations back into the time domain \cite{tzinis2022compute}. 

In the original implementation, SuDoRM-RF, \cite{tzinis2022compute} was used for single-channel speech separation. Here, we adapted the model to accommodate two-channel input. Furthermore, in several input configurations we added explicit spatial cues (IPD, see Section \ref{input_configuration}) as auxiliary input. In these cases, we concatenated IPD to the encoded latent representation of the mixture waveform using the concatenation function from the PyTorch library \cite{paszke2019pytorch} (Fig. \ref{img:AdaptedModel}B, bottom panel).

We utilised the following configurations for the SuDoRM-RF model: 512 input channels, 128 output channels and 16 U-blocks. The parameters were configured as follows: up-sampling depth = 4, encoder kernel size = 21 and number of encoder basis functions = 512. Further, the model utilises global layer normalization (gLN). These parameters are based on the original configuration \cite{tzinis2022compute}. Our version of SuDoRM-RF adapted for multi-channel speech separation and data is available here\footnote{
%Adapted Sudo RM-RF: 
\url{https://github.com/sayo20/Leveraging-Spatial-Cues-from-Cochlear-Implant-Microphones-to-Efficiently-Enhance-Speech-Separation-}}.

\section{Experimental procedures}

\subsection{Input configurations}
\label{input_configuration}
\noindent We trained the SuDoRM-RF model on various input configurations to quantify the impact of implicit and explicit spatial cues on speech separation performance in real-world listening scenes. As stated previously, we define \textit{implicit} spatial cues as cues that are inherently present in the data and can be learned by the model, but which are not manually extracted as an auxiliary input feature. Examples of such implicit spatial cues are inter-channel level differences (ILDs) and inter-channel phase differences (IPDs), which can be extracted from the comparison between two or more microphone channels. 

Three of the input configurations included in the present experiment (Fig. \ref{img:AdaptedModel}B) exclusively contained such \textit{implicit} spatial cues. Specifically, input samples consisting of a one-channel waveform contain implicit spatial cues in the form of level differences between the two talkers within the single channel. However, the implicit spatial cues in this input configuration are weak because no inter-channel comparison is possible. By contrast, both input samples consisting of two channels from a single CI (that is, unilateral, either from left or right ear) and input samples consisting of two channels of two different CIs (that is, bilateral, one on left and one on right ear) contain implicit spatial cues which can be extracted from an inter-channel comparison. However, implicit spatial cues are stronger in the two-channel, bilateral samples than in the two-channel, unilateral samples due to the position of the head in between the two channels. 

Further, in the present paper we refer to the IPDs added as auxilliary feature as \textit{explicit} spatial cues. To extract IPDs, we first converted waveforms to the time-frequency domain using the Short-Time Fourier Transform (STFT; hop size = 8 s, number of frequency bins = 512, and window length = 512 samples), and we matched the parameters with the frequency features as encoded by the encoders. We then calculated IPD similar to \cite{chen2018multi} as
\begin{equation}\label{eq:ipd}
    \textrm{IPD}(t,f) \coloneq \angle \left(\frac{y_{c_1}(t,f)}{y_{c_2}(t,f)}\right)
\end{equation}
where $y$ denotes the spectrogram representation of the two-talker mixture and $c_i$ denotes the channel. The IPD was mean normalized before being concatenated with the encoded mixed audio. 

Figure 3 B shows that we assessed three input configurations in the present experiment which contained also explicit spatial cues: one-channel unilateral waveform, two-channel unilateral waveforms, and the two-channel, bilateral waveforms (note that these configurations also include implicit spatial cues). For these input configurations, we calculated IPDs between the bilateral channels, that is, a channel from the left-ear CI and a channel from the right-ear CI, resulting in strong explicit spatial cues (Fig. \ref{img:AdaptedModel}B). 
In total, we trained the model on seven different input configurations (Fig. \ref{img:AdaptedModel}B and Table \ref{tab:train-test main result}).   
 
\subsection{Model training}
\label{model-training}
\noindent To train the network in an end-to-end manner, we utilised the scale-invariant signal-to-distortion ratio (SI-SDR) \cite{le2019sdr}:
\begin{equation}
\textrm{SI-SDR} \coloneq -10 \log_{10} \frac{||t_{target}||^2_2}{||e_{noise}||^2_2}
\end{equation}
Where  $e_{noise}$ is estimated target - true target ($t_{target}$).  When utilising real-world sound scenes, different versions of the target speech $t_{target}$ can be defined: A dry target (i.e. no reverberation similar to single-channel, monaural speech separation) or a reverberant target \cite{gu2020enhancing}). Here, we utilised dry targets for model training for two reasons. First, a dry speech waveform without reverberation is more beneficial for speech understanding in hearing impaired listeners and for cochlear implant users in particular. Second, dry targets are conventional in the large body of single-channel, monaural speech separation literature \cite{luo2019conv,luo2020dual}, making our results directly comparable to prior work.  

For every input configuration, the model was trained for 100 epochs with a batch size of 4. This batch size provided a good balance between computational efficiency and model performance. After 100 epochs, early stopping was applied (patience = 10, minimum delta = 0.1) to prevent overfitting. We utilised the network parameters corresponding to the best epoch to subsequently evaluate model performance on an independent test set. We used the Adam optimizer \cite{kingma2014adam} and a learning rate of $10^{-3}$ with a decay of 0.2 every 50 epochs. The model was implemented using the Pytorch framework \cite{paszke2019pytorch}. Finally, we used negative permutation-invariant training (PIT) \cite{yu2017permutation} to resolve the permutation problem.

\subsection{Online mixing} 
\label{sec:online mixing}
\noindent We implemented an augmented version of the mixing procedure described in WSJ0-2 mix\cite{hershey2016deep}. That is, talkers (either same gender or different gender) were mixed at a signal-to-noise ratio (SNR) ranging from 0 to 5 dB. However, to increase the variance in the training data we dynamically varied the fixed pairing of talkers when creating two-talker mixtures. Specifically, each training batch of four two-talker mixtures was randomly selected from the fixed samples in the WSJ0-2 mix\cite{hershey2016deep}. We then randomized the talkers among the four selected samples, while adhering to the original SNR distribution. 

All audio clips of two-talker mixtures were cut to a duration of four seconds. Shorter speech mixtures were zero-padded at the end or the beginning to the same length. The resulting mixtures were mean-variance normalized and downsampled to 8 kHz. 

\subsection{Evaluation metrics}
\noindent We employed the Scale-invariant Signal-to-Distortion Ratio (SI-SDR) \cite{le2019sdr} and its improvement variant (SI-SDRi) as evaluation metrics to assess the performance of the model. The SI-SDRi quantifies the increase in SI-SDR in the cleaned speech waveforms in comparison to the initial two-talker mixture. Although the model was trained on the SI-SDR loss (see Section \ref{model-training}), the SI-SDRi is more informative for comparing speech separation performance across different input configurations due to the differences in baseline SI-SDR between input configurations as a result of the spatialisation and reverberation. We calculated both distortion metrics using the Asteroid framework \cite{Pariente2020Asteroid}. Additionally, we measured the perceptual quality of the cleaned speech segments using the Short-Time Objective Intelligibility (STOI, range [0,1])\cite{taal2011algorithm} and Perceptual Evaluation Speech Quality (PESQ, range [-0.5, 4]) \cite{rix2001perceptual} metrics. Both perceptual metrics were derived using the Torchaudio Toolbox \cite{yang2022torchaudio}. Note that by utilising dry targets (Section \ref{input_configuration}), separation performance metrics are also affected by remaining reverberation in the clean speech waveforms.

\section{Results}
\noindent The present work aimed to quantify the impact of implicit and explicit spatial cues captured by CI microphones on speech separation in real-world listening scenes. To this end, we assessed and compared speech performance for seven different input configurations, which vary in the presence and strength of implicit and explicit spatial cues (Fig. \ref{img:AdaptedModel}B and Table \ref{tab:train-test main result}).

\begin{table*}[htpb]
\centering
\caption{Speech separation performance for non-spatial, dry speech mixtures and for spatial, reverberant speech mixtures.}
\begin{tabular}{c|c|c|c|c|c|c|c}
\hline
\textbf{Acoustic scene} & \textbf{Input features} & \textbf{Spatial information} & \textbf{\#Params} & \textbf{SI-SDR$\uparrow$} & \textbf{SI-SDRi$\uparrow$} & \textbf{STOI$\uparrow$} & \textbf{PESQ$\uparrow$} \\
\hline
Non-spatial, dry & One-channel waveform & None & 2.6M & 12.96 & 12.96 & 0.89 & 2.97 \\
\hline
Spatial, reverberant & One-channel waveform & Implicit & 2.6M & 2.66 & 27.01 & 0.70 & 1.81 \\
Spatial, reverberant & Two-channel, unilateral waveform & Implicit & 2.6M & 2.74 & 27.99 & 0.73 & 1.88 \\
Spatial, reverberant & Two-channel, bilateral waveform & Implicit & 2.6M & \textbf{3.69} & \textbf{28.05} & \textbf{0.77} & \textbf{2.02} \\
\hline
Spatial, reverberant & One-channel waveform + IPD & Implicit + Explicit & 3.8M & 2.74 & 27.09 & 0.74 & 1.92 \\
Spatial, reverberant & Two-channel, unilateral waveform + IPD & Implicit + Explicit & 3.8M & 4.16 & \textbf{29.41} & 0.78 & 2.09 \\
Spatial, reverberant & Two-channel, bilateral waveform + IPD & Implicit + Explicit & 3.8M & \textbf{4.36} & 28.72 & \textbf{0.78} & \textbf{2.12} \\
\hline
\end{tabular}
\label{tab:train-test main result}
\end{table*}

\subsection{Baseline: Dry, non-spatial scenes}
\noindent To establish a baseline, we first trained and evaluated the SudoRM-RF model on one-channel, dry and non-spatial two-talker mixtures (i.e., the original WSJ0-2mix dataset \cite{hershey2016deep}). As outlined in Table \ref{tab:train-test main result} (row 1), the model obtained a SI-SDRi of 12.96 dB for this dataset, indicating that the model accurately separated two concurrent speech streams. Although other, larger speech separation models outperform the current SuDoRM-RF implementation on the WSJ0-2mix dataset\cite{luo2019conv,subakan2021attention}, it should be noted that we purposefully selected a small and efficient model that can potentially be deployed on a CI. Moreover, we did not perform any type of data pre-processing (for example, silence removal) to ensure that sound scenes maintain their natural characteristics. 

\subsection{Speech separation in real-world acoustic scenes}
\noindent We quantified to what extent speech separation performance of the SuDoRM-RF model deteriorated when the model was trained on one-channel, real-world speech mixtures (Table \ref{tab:train-test main result}, row 1 and 2). Results show that the overall quality of the resulting separated speech waveforms was substantially lower when the model was trained on one-channel, real-world two-talker mixtures (SI-SDR = 2.66, STOI = 0.70, PESQ = 1.81) than when the model was trained on one-channel, non-spatial, dry two-talker mixtures (SI-SDR = 12.96, STOI = 0.89 and PESQ = 2.97; Table \ref{tab:train-test main result}). However, the SI-SDRi score was substantially higher for the model trained on real-world scenes (27.01 dB) than for the model trained on non-spatial, dry scenes (12.96 dB). This discrepancy between SI-SDRi and SI-SDR can be explained by the difference in baseline SI-SDR of the two input configurations: the real-world two-talker mixtures had a baseline SI-SDR of -24.26 dB, while the non-spatial, dry two-talker mixtures had a SI-SDR baseline of 0 dB. Importantly, the observed decline of 79.4\% in SI-SDR demonstrates that a model which performs well on non-spatial, dry two-talker mixtures cannot be generalized directly to real-world, spatial and reverberant acoustic scenes without a significant drop in performance. 

\subsection{Impact of implicit and explicit spatial cues on speech separation}
\label{use-of-spatial}
\noindent Next, we investigated to what extent speech separation performance in real-world listening scenes improves when the model has access to implicit and explicit spatial cues. 

Table \ref{tab:train-test main result} (rows 2 - 4) shows that speech separation performance improved when model input contained strong implicit spatial cues. Specifically, separation performance was substantially higher when the model was trained on two-channel, bilateral waveforms than when the model was trained on one-channel, unilateral waveforms with weak implicit spatial cues (+38.7\% SI-SDR, +10\% STOI, +11.6\% PESQ) or two-channel, unilateral waveforms with intermediate spatial cues (+3.0\% SI-SDR, +4.3\% STOI, +3.9\% PESQ).

Table \ref{tab:train-test main result} shows that adding explicit spatial cues (i.e., IPDs) to the model input improved performance for all input configurations (rows 5 - 7). However, the extent to which explicit spatial cues improved speech separation performance varied. Adding IPDs as an auxiliary feature had the strongest impact on two-channel, unilateral waveforms (+51.8\% SI-SDR, +6.8\% STOI and +11.1\% PESQ), followed by two-channel, bilateral waveforms (+18.2\% SI-SDR, +1.3\% STOI and +5.0\% PESQ) and finally one-channel, unilateral waveforms (+3.0\% SI-SDR, +5.7\% STOI and +6.1\% PESQ).

\begin{figure}[h]
    \centering
    \includegraphics[width=\linewidth]{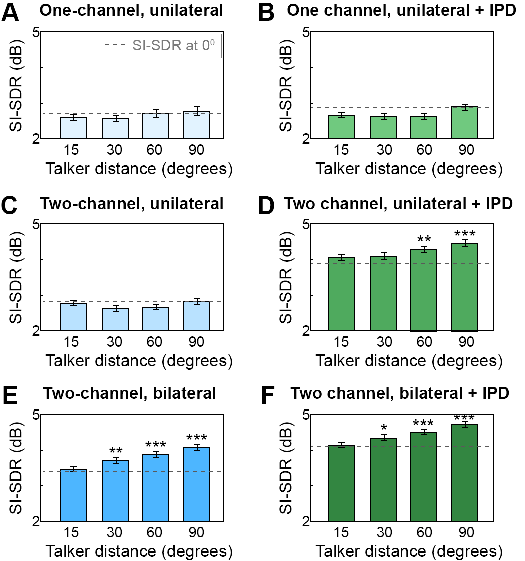}
    \caption{Speech separation performance as a function of spatial separation angle between talkers. Each panel depicts the results for one input configuration (Fig. 3B, Table \ref{tab:train-test main result}). Colors indicate implicit (blue) and combined implicit and explicit (green) spatial information load, with higher saturation indicating higher spatial information load (similar to Fig. 3B). Bars represent average SI-SDR, error bars reflect standard error of the mean. As a reference, the gray dashed line indicates the average SI-SDR at a separation angle of 0°, that is, no spatial distance. Asterisks reflect a statistically significant difference in SI-SDR at that relevant separation angle and the 0° reference: * = \textit{p} $<$ 0.05, ** = \textit{p} $<$ 0.01.}
\label{img:spatialcue}
\vspace{-5pt} % this sets the distance between the figure and the text
\end{figure}

\begin{table}[t]
\caption{Spatial distance affects speech separation. Kruskal-Wallis H tests, corrected for multiple comparisons with the False Discovery Rate (FDR, \cite{benjamini1995controlling}).}
\label{tab:angle_effect}
\begin{center}
\begin{adjustbox}{width=\columnwidth,center}
\begin{tabular}{l|c|c}
\toprule
\textbf{Input configuration} & \textbf{$\chi$$^2$}(4) & \textbf{Significance} \\
\toprule
One-channel waveform & 5.07 & 0.31 \\
Two-channel, unilateral waveform & 4.82 & 0.31 \\
Two-channel, bilateral waveform & 55.6 & 1.4e-8***\\
\midrule
One-channel waveform + IPD & 11.9 & 0.03*\\
Two-channel, unilateral waveform + IPD & 28.8 & 1.7e-5***\\
Two-channel, bilateral waveform + IPD & 43.1 & 3.0e-8***\\
\bottomrule
\end{tabular}
\end{adjustbox}
\end{center}
\vspace{-10pt} % this sets the distance between the table and the text
\end{table}

\subsection{Utilisation of spatial cues}
\noindent To evaluate how the model utilises implicit and explicit spatial cues when performing speech separation in real-world sound scenes, we assessed whether spatial separation affected speech separation performance (Kruskal-Wallis H tests, FDR corrected for multiple comparisons). The results revealed an effect of spatial separation angle on SI-SDR for all input configurations containing explicit spatial cues as well as for input configuration with strong implicit spatial cues (two-channel, bilateral waveforms), but not for input configurations with weak or intermediate implicit spatial cues (Table \ref{tab:angle_effect}). 

Subsequent post-hoc pairwise comparisons for the input configurations which exhibited a significant effect of spatial separation angle, demonstrated that speech separation performance improved with increasing angles of spatial separation between talkers (Table II, Fig. \ref{img:spatialcue}D, \ref{img:spatialcue}E, \ref{img:spatialcue}F). By contrast, input configurations that contained weak or intermediate implicit spatial cues exhibited uniform speech separation performance across spatial separation angles (one-channel, unilateral waveforms or two-channel, unilateral waveforms; Fig. \ref{img:spatialcue}A, \ref{img:spatialcue}C). Finally, in agreement with the weak effect of spatial separation angle on speech separation performance for the input configuration consisting of one-channel, unilateral waveforms +IPD (Table \ref{tab:angle_effect}), post-hoc pairwise comparisons did not reveal any significant performance differences as a function of spatial separation angle for this input configuration (Fig. \ref{img:spatialcue}B)These results illustrate that the model leveraged spatial cues (both implicit and explicit) in the input to improve speech separation performance in real-world acoustic scenes. 

Strikingly, our results show that the presence of spatial cues in input data also improved speech separation performance for spatially overlapping talkers (compare dashed lines between input configurations with and without spatial cues in Fig. \ref{img:spatialcue}). 

\subsection{Spectral and spatial cues for speech separation interact}
\noindent A speech separation model trained on conventional non-spatial, dry acoustic scenes uses spectral differences between talkers to separate the speech streams \cite{qian2018past}. Here, we hypothesized that for a model performing speech separation on real-world acoustic scenes, the presence of spatial cues in the data is especially beneficial for speech separation when talkers' voices are spectrally similar - that is, when spectral cues are weak. To evaluate this hypothesis, we assessed the effect of spatial separation angle on speech separation performance as a function of talker gender pairing. Figure 5 and Table III show that the effect of spatial separation angle on speech separation performance was larger for speech mixtures consisting of two female talkers or two male talkers than for mixtures consisting of a male and a female talker (Kruskall-Wallis H tests, FDR corrected). For example, for the input configuration consisting of two-channel, unilateral waveforms + IPD, speech separation performance was 23.0\% higher at large spatial separation angles than at small separation angles for mixture consisting of two male talkers,  12.0\% for mixtures consisting of two female talkers, and 4.9 \% for mixtures consisting of one male and one female talker. Taken together, these findings demonstrate that the presence of spatial cues in real-world scenes is especially beneficial for speech separation when talkers have similar spectral profiles.

\begin{figure}[h]
    \centering
    \includegraphics[width=\linewidth]{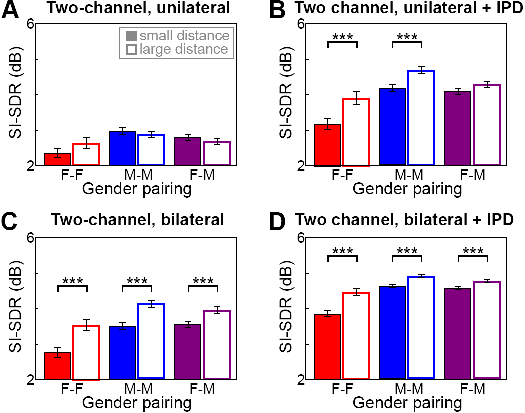}
    \caption{SI-SDR as a function of talker gender pairing and spatial distance for various input configurations. Bars represent average SI-SDR at small spatial distances (0\textdegree{} and 15\textdegree{}, filled bars) or and large spatial distances (60\textdegree{} and 90\textdegree{}, open bars) across the different gender pairings (red = two female talkers; blue = two male talkers; purple = one male and one female talker). Asterisks reflect a statistically significant difference (Kruskal-Wallis H tests, FDR corrected): *** = \textit{p} $<$ 0.001; ** = \textit{p} $<$ 0.01.}
\label{img:gender-sep}
\end{figure}

\begin{table*}[!t]
\caption{Speech separation performance as a function of gender pairing and spatial distance.}
\label{tab:gender_pairing}
\begin{adjustbox}{width=0.95\textwidth,center}
\begin{tabular}{lcccccc}
\toprule
\multirow{2}{*}{\textbf{Input configuration}} & \textbf{Gender pairing} & \textbf{Small distance} & \textbf{Large distance} & \textbf{Improvement} & \textbf{$\chi$$^2$(1)} & \textbf{Significance} \\
& & SI-SDR (dB) & SI-SDR (dB) & (\%) & & \\
\midrule
\multirow{3}{*}{Two-channel, bilateral} & F-F & 2.77 & 3.53& 27.42 & 14.30 & 2.0e-4*** \\
& M-M & 3.50 & 4.13 & 18.00 & 24.93 & 1.8e-6*** \\
& F-M & 3.55 & 3.96 & 11.55 & 13.21 & 0.002** \\
\midrule
\multirow{3}{*}{Two-channel, unilateral + IPD} & F-F & 3.17 & 3.90 & 23.02 & 11.28 & 0.001** \\
& M-M & 4.17 & 4.67 & 12.00 & 12.98 & 0.001** \\
& F-M & 4.08 & 4.28 & 4.90 & 3.04 & 0.1 \\
\midrule
\multirow{3}{*}{Two-channel, bilateral + IPD} & F-F &3.07 & 4.08 &32.90 & 22.54  & 6.2e-6*** \\
& M-M & 4.36 & 4.82 & 10.55 & 12.80 & 6.0e-4*** \\
& F-M & 4.26 & 4.60 &7.98 &  9.74& 0.002** \\
\bottomrule
\end{tabular}
\end{adjustbox}
\vspace{0pt}
\end{table*}

\section{Discussion}
\noindent This study investigated the efficacy of utilising spatial cues derived from cochlear implant microphones to improve speech separation in real-world, spatial and reverberant scenes in an efficient manner using the time-domain speech separation model SudoRM-RF\cite{tzinis2022compute}.  

Similar to other studies performing speech separation on spatial and reverberant speech mixtures \cite{maciejewski2020whamr,zhao2023mossformer2,han2020real}, we observed a notable decline in the model's performance when trained on real-world acoustic scenes in comparison to when trained on non-spatial, dry scenes. These findings confirm that a model developed for speech separation of non-spatial, dry two-talker mixtures does not generalize robustly to spatial, reverberant two-talker mixtures. This highlights the importance of employing ecologically valid data for the development and optimization of speech separation models for real-world applications. 

Unlike existing approaches which aim to improve speech separation by adding spatial cues as an auxilliary feature and thereby unavoidably reduce model efficiency (e.g. \cite{gu2020temporal}), our approach demonstrates that a time-domain model can learn to extract and utilise the naturally available spatial cues that are present in the signals captured by CI microphones to efficiently optimise speech separation. This aligns with how individuals with normal hearing utilise spatial cues for auditory scenes analysis \cite{van2019cortical}. Moreover, we showed that spatial cues are especially beneficial for speech mixtures consisting of talkers with similar voices, such as same-gender speech mixtures. These results emphasize the potential of utilizing naturally available spatial information captured by microphones on assistive hearing devices for speech separation, to increase computational efficiency for such real-world applications.

Furthermore, we show that adding explicit spatial cues in the form of IPDs to time-domain data has a larger impact when the data consists of two channels (either unilateral or bilateral) than when they consist of a single channel.  As the addition of IPDs as auxilliary feature affects model latency (due to the STFT required to calculate IPD) as well as model size (+46\% parameters, see Table \ref{tab:train-test main result}), elucidating in which situations implicit spatial situations are sufficient and in which situations adding explicit spatial cues improves speech separation is highly relevant for applications such as CIs, which require real-time, computationally efficient models.  

Strikingly, we found that the availability of spatial cues improved speech separation also for speech mixtures consisting of two spatially overlapping talkers for which spatial cues are comparable. It is not directly clear what causes this upward shift. We hypothesize that during the spatial learning that occurs when the speech mixtures contain spatial cues (as demonstrated by our results), the model does not only learn the frequency-dependent nature of spatial cues but also the frequency-dependent nature of reverberation characteristics \cite{lollmann2011estimation} from the samples consisting of spatially separated talkers. Conceivably, the learned frequency-specific reverberation characteristics reinforce the spectral cues for separating speech of spatially overlapping talkers. Importantly, these findings indicate that providing front-end speech processing technology in a CI or other assistive hearing device with multi-channel input - and thereby the naturally available spatial cues - will improve the separation of speech both for spatially separated talkers and for talkers that are close to each other. 

While most existing speech separation approaches trained on real-world data use a reverberant version of the separated speech streams as target \cite{gu2020enhancing,han2020real}, we used a dry version of the speech stream as target. Our rationale was that a dry target directs the model towards simultaneous speech separation and speech dereverberation. De-reverberation is advantageous for the present use case of CIs, as the presence of reverberation in speech waveforms reduces speech intelligibility especially for hearing impaired listeners \cite{nabelek1974monaural,xia2018effects}. However, the separated speech streams still contained substantial reverberation. Therefore, more research into strategies for integrating speech dereverberation approaches with speech separation approaches \cite{doclo2015multichannel,pfeifenberger2022blind,ni2021wpd++} is needed. 

\section{Conclusion}
\noindent This study explored the potential of leveraging spatial cues derived from cochlear implant microphones for efficient speech separation in real-world acoustic scenes. Our results highlight that ecologically valid data is crucial for the development of speech separation technology for real-world applications such as front-end speech processing in a CI or other assistive hearing device. Moreover, our findings demonstrate that strong implicit spatial cues efficiently boost speech separation accuracy in real-world listening scenes and that adding explicit spatial cues as auxiliary feature boosts speech separation accuracy when implicit spatial cues are only weakly present in speech mixtures. These insights pave the way for the development of more efficient speech separation approaches for listeners using CIs or other assistive hearing devices in everyday, noisy listening situations.

\section{Acknowledgement}
\noindent This work is part of the INTENSE consortium, which has received funding from the NWO Cross-over Grant No. 17619. We thank Advanced Bionics for providing the CI-HRTFs.
% -------------------------------------------------------------------------
% \bibliographystyle{IEEEbib}
\bibliographystyle{ieeetr}
\bibliography{Main}

\end{document}